\documentclass[twocolumn,prb,showpacs,superscriptaddress,preprintnumbers,amsmath,amssymb]{revtex4}

\usepackage{graphicx}

\begin{document}
\title{Valence, spin, and orbital state of the Co ions in the one-dimensional
Ca$_{3}$Co$_{2}$O$_{6}$: an x-ray absorption and magnetic circular
dichroism study}

\author{T. Burnus}
  \affiliation{II. Physikalisches Institut, Universit{\"a}t zu K{\"o}ln,
   Z{\"u}lpicher Str. 77, D-50937 K{\"o}ln, Germany}
\author{Z. Hu}
  \affiliation{II. Physikalisches Institut, Universit{\"a}t zu K{\"o}ln,
   Z{\"u}lpicher Str. 77, D-50937 K{\"o}ln, Germany}
\author{M. W. Haverkort}
  \affiliation{II. Physikalisches Institut, Universit{\"a}t zu K{\"o}ln,
   Z{\"u}lpicher Str. 77, D-50937 K{\"o}ln, Germany}
\author{J. C. Cezar}
  \affiliation{European Synchrotron Radiation Facility, Bo\^\i{}te Postale 220,
   Grenoble 38043, France}
\author{D. Flahaut}
  \affiliation{Laboratoire CRISMAT, UMR 6508, ENSICAEN/CNRS, 
  Universit\'e de Caen, 6, Boulevard du Mar\'echal Juin, 14050 Caen Cedex, France}
\author{V. Hardy}
  \affiliation{Laboratoire CRISMAT, UMR 6508, Institut des Sciences de la Mati\`ere et du Rayonnement,
  Universit\'e de Caen, 6, Boulevard du Mar\'echal Juin, 14050 Caen Cedex, France}
\author{A. Maignan}
  \affiliation{Laboratoire CRISMAT, UMR 6508, Institut des Sciences de la Mati\`ere et du Rayonnement,
  Universit\'e de Caen, 6, Boulevard du Mar\'echal Juin, 14050 Caen Cedex, France}
\author{N. B. Brookes}
  \affiliation{European Synchrotron Radiation Facility, Bo\^\i{}te Postale 220,
   Grenoble 38043, France}
\author{A. Tanaka}
  \affiliation{Department of Quantum Matter, ADSM, Hiroshima University, Higashi-Hiroshima 739-8530, Japan}
\author{H. H. Hsieh}
  \affiliation{Chung Cheng Institute of Technology, National Defense University, Taoyuan 335, Taiwan}
\author{H.-J. Lin}
  \affiliation{National Synchrotron Radiation Research Center, 101 Hsin-Ann Road, Hsinchu 30077, Taiwan}
\author{C. T. Chen}
  \affiliation{National Synchrotron Radiation Research Center, 101 Hsin-Ann Road, Hsinchu 30077, Taiwan}
\author{L. H. Tjeng}
  \affiliation{II. Physikalisches Institut, Universit{\"a}t zu K{\"o}ln,
   Z{\"u}lpicher Str. 77, D-50937 K{\"o}ln, Germany}

\date{\today}

\begin{abstract}

We have investigated the valence, spin, and orbital state of the
Co ions in the one-dimensional cobaltate Ca$_3$Co$_2$O$_6$ using
x-ray absorption and x-ray magnetic circular dichroism at the
Co-$L_{2,3}$ edges. The Co ions at both the octahedral Co$_{\rm
oct}$ and trigonal Co$_{\rm trig}$ sites are found to be in a 3+
state. From the analysis of the dichroism we established a
low-spin state for the Co$_{\rm oct}$ and a high-spin state with an
anomalously large orbital moment of $1.7\mu_B$ at the
Co$^{3+}_{\rm trig}$ ions. This large orbital moment along the
$c$-axis chain and the unusually large magnetocrystalline
anisotropy can be traced back to the double occupancy of the
$d_{2}$ orbital in trigonal crystal field.

\end{abstract}

\maketitle

The one-dimensional Ca$_3$Co$_2$O$_6$ has attracted great
interest in recent years due to the observation of the stair-step
jumps in the magnetization at regular intervals of the applied
magnetic field.
\cite{Aasland:1997,Kageyama:1997,Kageyama:1997b,Maignan:2000,Kudasov:2006,Hardy:2004a,Flahaut:2004,Hardy:2004,Petrenko2005}
The rhombohedral structure of this compound consists of
[Co$_2$O$_6$]$_\infty$ chains running along the $c$ axis of the
hexagonal unit cell.\cite{Fjellvaag:1996} In each chain, CoO$_6$
octahedra alternate with CoO$_6$ trigonal prisms. The magnetism
is Ising-like and directed along the Co chains with large
magnetic moments of 4.8 $\mu_B$ per formula
 unit.\cite{Hardy:2003,Maignan:2004} The intra-chain coupling is
ferromagnetic with a spin-freeze at $T_{\rm SF}=7$~K and the
chains couple antiferromagneticly with a N\'{e}el temperature of
$T_N = 25\rm~K$.\cite{Hardy:2003}

Based on density-functional-theory calculations, Vidya \textit{et
al.} claimed a low-spin (LS) Co$^{4+}_{\rm oct}$ and a high-spin
(HS) Co$^{2+}_{\rm trig}$ state for Ca$_3$Co$_2$O$_6$.
\cite{Vidya:2003} However, other ex\-pe\-ri\-men\-tal and theoretical
works have proposed a Co$^{3+}$ valence state at both the
octahedral and trigonal Co sites, with the Co$_{\rm oct}$ LS
($S=0$) and the Co$_{\rm trig}$ HS ($S=2$)
state.\cite{Sampathkumaran:2004,Takubo:2005,Eyert:2004,Whangbo:2003,Wu:2005}
The Ising character of the magnetism is also subject of
discussion. Dai and Whangbo \textit{et al.} found from their band
structure calculations that the spin-orbit-inactive $d_0$ orbital
lies lowest of all Co$_{\rm trig}$ crystal-field
levels,\cite{Whangbo:2003,Dai:2005} and had to invoke excited states in
their attempt to explain the Ising magnetism. Wu
\textit{et al.},\cite{Wu:2005} on the other hand, calculated that it is the
spin-orbit-active $d_2$ orbital which lies lowest, giving a very
different picture for the Ising magnetism.

In this paper we have applied soft-x-ray absorption spectroscopy
(XAS) and magnetic circular dichroism (XMCD) at the Co-$L_{2,3}$
edges to resolve the Co valence, spin and orbital state issue in
Ca$_3$Co$_2$O$_6$. We have also carried out detailed
configuration-interaction cluster calculations to analyze the
spectra. We found that the Co ions are all in the 3+ state, that
the Co$_{\rm oct}$ are non-magnetic, and that the Co$^{3+}_{\rm
trig}$ carry a $1.7\mu_B$ orbital moment. We clarify the orbital
occupation issue and the origin of the Ising magnetism.

The
Ca$_3$Co$_2$O$_6$ single crystals were grown by heating a
mixture of Ca$_3$Co$_4$O$_9$ and K$_2$CO$_3$
in a weight ratio of 1:7 at 950
$^{\circ}$C for 50 h in an alumina
crucible in air. The cooling was performed in two steps,
first down to 930 $^{\circ}$C at a rate of 10 $^{\circ}$C/h and
then down to room temperature at 100
$^{\circ}$C/h.\cite{Hardy:2004} The Co-$L_{2,3}$ XAS
spectra of
Ca$_3$Co$_2$O$_6$ and of CoO and EuCoO$_3$ (Ref. \onlinecite{Hu:2004}) as references were
recorded at the Dragon beamline of the National Synchrotron
Radiation Research Center (NSRRC) in Taiwan with an energy
resolution of 0.3 eV. The first sharp peak at 777 eV of
photon energy at the Co-$L_3$ edge of CoO was used for energy
calibration, which enabled us to achieve better than 0.05 eV
accuracy in the relative energy alignment. The XMCD spectra were collected at the ID08 beamline of
the European Synchrotron Radiation Facility (ESRF) in Grenoble
with a resolution of 0.25 eV and a degree of circular polarization
close to $100\%$ in a magnetic field of 5.5 Tesla with the sample
kept at approximately 15--20~K, using a dedicated superconducting magnet
setup from Oxford Instruments.
The Poynting vector of the photons and the magnetic field were
both parallel to the $c$ axis. The single-domain Ca$_3$Co$_2$O$_6$
crystal used for the XMCD experiment was needle-shaped with a
dimension of $0.2\times0.2\times10$~mm$^3$ for $a\times b\times
c$. Clean sample areas were obtained by cleaving the crystals
\textit{in situ} in chambers with base pressures in the low
10$^{-10}$ mbar range. The Co $L_{2,3}$ spectra were recorded
using the total-electron-yield method (TEY).

\begin{figure}
    \includegraphics[width=0.5\textwidth]{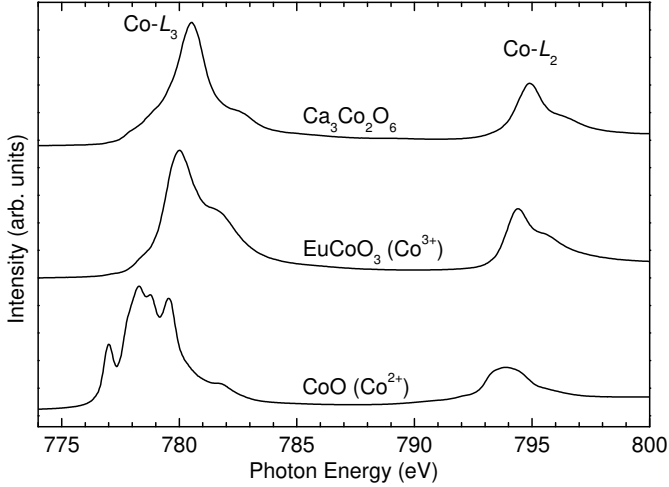}
    \caption{Co-$L_{2,3}$ XAS spectra of Ca$_{3}$Co$_{2}$O$_{6}$,
     CoO, and EuCoO$_3$.}
    \label{XAS}
\end{figure}

Fig.~\ref{XAS} shows the Co-$L_{2,3}$ XAS spectrum of
Ca$_{3}$Co$_{2}$O$_{6}$ together with that of CoO and EuCoO$_3$.
CoO serves here as a $2+$ reference and EuCoO$_3$ as a LS $3+$
reference.\cite{Hu:2004} One can see first of all that the $2+$ spectrum (CoO)
contains peaks which are 2 or more eV lower in energy than the
main peak of the $3+$ spectrum (EuCoO$_3$). It is well
known that x-ray absorption spectra at the transition-metal
$L_{2,3}$ edges are highly sensitive to the valence state. An increase of the valence of the metal ion by
one results in a shift of the $L_{2,3}$ XAS spectra to higher
energies by 1~eV or more.\cite{Mitra2003} This shift is due to a final
state effect in the x-ray absorption process. The energy difference between a $3d^n$ ($3d^7$ for Co$^{2+}$) and a $3d^{n-1}$ ($3d^6$ for Co$^{3+}$) configuration is $\Delta E=E(2p^63d^{n-1}\to2p^53d^{n})-E(2p^63d^n\to2p^53d^{n+1})\approx U_{\underline pd}-U_{dd}\approx 1$--2~eV, where $U_{dd}$ is the Coulomb repulsion energy between two $3d$ electrons and $U_{\underline pd}$ the one between a $3d$ electron and the $2p$ core hole.

One can observe from Fig.~\ref{XAS} that the Ca$_3$Co$_2$O$_6$ spectrum has no features at the low-energy side, which otherwise could have indicated the presence of Co$^{2+}$ species like in CoO. Instead, the spectrum has a much closer resemblance to that of Co$^{3+}$, like in EuCoO$_3$.
 This means that one
can safely rule out the Co$^{2+}$/Co$^{4+}$
scenario.\cite{Vidya:2003} In other words, the XAS experiment
reveals unambiguously that both the Co$_{\rm oct}$ and the
Co$_{\rm trig}$ ions are in the $3+$ valence state. This result
supports the analysis of the Co $2p$ core-level x-ray
photoemission spectra\cite{Takubo:2005} and band-structure
calculations\cite{Eyert:2004,Wu:2005}.

\begin{figure}
    \includegraphics[width=0.5\textwidth]{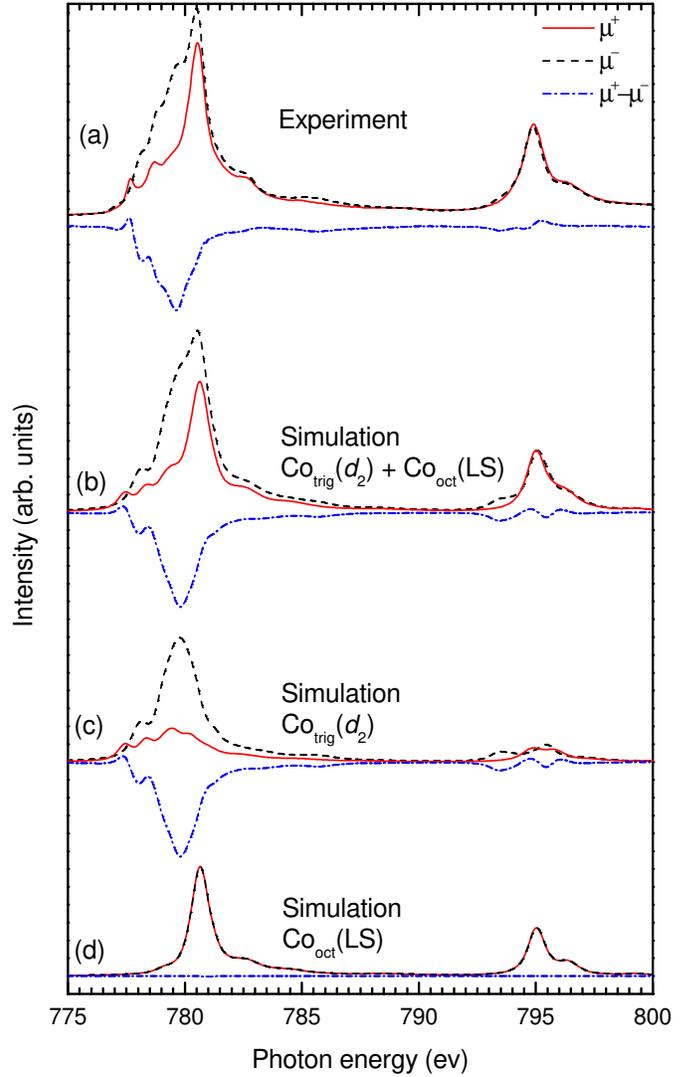}
    \caption{(Color online) (a) Measured soft-x-ray absorption
     spectra with parallel ($\mu^+$, red solid line) and
     antiparallel ($\mu^-$, black dashed line) alignment between
     photon spin and magnetic field, together with the difference
     spectrum ($\mu^+-\mu^-$, blue dash-dotted);
     (b) Simulated sum spectra assuming a doubly occupied $d_2$
     orbital for the Co$_{\rm trig}$ and low-spin (LS)
     Co$_{\rm oct}$ ions;
     (c) and (d) Contribution of the Co$_{\rm trig}$ and
     Co$_{\rm oct}$ ions to the simulated sum spectra.}
    \label{XMCD1}
\end{figure}

In order to resolve the spin-state issue, we now resort to XMCD.
The top part (a) of Fig.~\ref{XMCD1} depicts Co-$L_{2,3}$ XAS
spectra taken with circularly polarized light with the photon spin
parallel (red solid line, $\mu^{+}$) and antiparallel (black
dashed line, $\mu^{-}$) aligned to the magnetic field. The
quantization axis $z$ has been chosen to be parallel to the $c$
axis, which is the easy magnetization axis\cite{Kageyama:1997b}.
One can clearly observe large differences between the spectra
using these two alignments. The difference spectrum, i.e. the XMCD
spectrum, is also shown (blue dash-dotted line,
$\mu^{+}-\mu^{-}$). Using the well-known XMCD sum rule developed
by Thole \textit{et al.},\cite{Thole:1992}

\begin{equation}\label{sumruleLz}
  L_z = \frac{4}{3}
 \frac{\int [\mu^+(E) - \mu^-(E)]\,{\rm d}E}
      {\int[\mu^+(E)+\mu^-(E)]\,{\rm d}E} (10-N_e),
\end{equation}

\noindent where $N_e$ is the number of the Co $d$
electrons, we can extract directly the orbital ($L_z$)
contribution to the magnetic moment without the need to do
detailed modeling. Assuming an average Co $3d$ occupation number
of about $N_e=6.5$ electrons
(estimated for a HS Co$^{3+}$ oxide\cite{Saitoh1997} and
also to be justified below by cluster calculations) we find a
value of $1.2\mu_B$ for the $L_z$, which is a very large number
indicating that the ground state is strongly spin-orbit active.

To extract more detailed information concerning the spin and
orbital states from the Co $L_{2,3}$ XAS spectra, we have
carried out simulations for the XMCD spectra
 using the well-proved
configuration-interaction (CI) cluster
model.\cite{Tanaka94,deGroot94,Thole97} The method uses for each
Co site a CoO$_6$ cluster which includes the full atomic multiplet
theory and the local effects of the solid. It accounts for the
intra-atomic $3d$--$3d$ and $2p$--$3d$ Coulomb interactions, the
atomic $2p$ and $3d$ spin-orbit couplings, the O $2p$--Co $3d$
hybridization, and the proper local crystal-field parameters.
In the configuration-interaction cluster calculation we describe the
ground state by the configurations

\begin{equation}
\Psi=\alpha_6|d^6\rangle+\alpha_7|d^7\underline
L\rangle+\alpha_8|d^8\underline L^2\rangle+\alpha_9|d^9\underline
L^3\rangle,
\end{equation}

\noindent where $\underline L$ denotes a ligand hole and
$\sum_{i=6}^9\alpha_i^2=1$.\cite{Saitoh1995,Hu:1998} The
transition-metal electron occupation is then given by

\begin{equation}
N_e=6\alpha_6^2 + 7\alpha_7^2 + 8\alpha_8^2+9\alpha_9^2.
\end{equation}

\noindent The
simulations have been carried out using the program XTLS
8.3.\cite{Tanaka94}

In octahedral symmetry the $3d$ orbitals split up in the
well-known $e_{g}$ and $t_{2g}$ levels; the splitting is given by
$10Dq$. In a trigonal prismatic environment, however, it is found
that the $x^{2}-y^{2}$ is degenerate with the $xy$, and the $yz$
with $zx$ orbital.\cite{Whangbo:2003,Wu:2005} In the presence of
the spin-orbit coupling, it is then better to use the complex
orbitals $d_0$, $d_{2}$/$d_{-2}$, and $d_{1}$/$d_{-1}$. Band
structure calculations indicate that the $d_1$/$d_{-1}$ band is
split off from the $d_0$ and $d_2$/$d_{-2}$ bands by about 1 eV,
while the $d_0$ and $d_2$/$d_{-2}$ bands are almost
degenerate.\cite{Whangbo:2003,Wu:2005} Critical for the magnetism
and for the line shape of the simulated spectra are the crystal
field parameter $10Dq$ for the Co$_{\rm oct}$O$_6$ cluster and
the crystal field parameter for the splitting between the nearly
degenerate $d_0$ and d$_2$ orbitals of the Co$_{\rm trig}$O$_6$
cluster. $10Dq$ needs to be critically tuned since this is set to
determine whether Co$_{\rm oct}$--O$_6$ cluster is in the LS or HS
state\cite{Goodenough:1971}. The trigonal prism crystal field has
also to be fine tuned since it determines whether the $d_2$ or the
$d_0$ lies lowest, and thus the magnitude of the orbital moment
and strength of the magnetocrystalline anisotropy as we will show
below. Tuning of these parameters will be done by establishing
which of the simulated spectra reproduce the experimentally
observed ones.

As a starting point, we make the assumption that the Co$_{\rm
oct}$ ion is in the LS state, based on the observation that the
average Co$_{\rm oct}$-O bond length of 1.916~\AA\ in this
material\cite{Fjellvaag:1996} is shorter than the 1.925~\AA in LaCoO$_3$ at 5~K,
 which is known to be LS\cite{Radaelli2002}. With such a short length, the
Co ion is subjected to a large enough $10Dq$, sufficient to
stabilize the non-magnetic LS
state\cite{Goodenough:1971,Radaelli2002} for Co$_{\rm
oct}^{3+}$ ions in Ca$_3$Co$_2$O$_6$.  With
this starting point, the magnetism and the XMCD signal have to
originate from the Co$_{\rm trig}$ ions. This is quite plausible
since with the extremely large Co$_{\rm trig}$--O average bond
length of 2.062~\AA,\cite{Fjellvaag:1996} which is much
larger than 1.961~\AA\ for LaCoO$_3$ at
1000~K,\cite{Radaelli2002} one can expect that the crystal field
is small enough to stabilize the HS state.\cite{Goodenough:1971}
Based on the observation that the orbital contribution to the
magnetic moment is extremely large, we take the $d_2$
\textit{ansatz} for the Co$_{\rm trig}$ and not the $d_0$.
In the CI calculation, the parameters for the multipole
part of the Coulomb interactions were given by the Hartree-Fock
values, while the monopole parts ($U_{dd}$, $U_{pd}$) were
estimated from photoemission experiments on Co$^{3+}$
materials\cite{Bocquet96}. The one-electron parameters such as the
O $2p$--Co $3d$ charge-transfer energies and integrals were
estimated from band-structure
results.\cite{Whangbo:2003,Dai:2005,Wu:2005} The charge-transfer
energy is given by $\Delta=E(d^6)-E(d^7 \underline L)=1.5$~eV, the
$d$--$d$ Coulomb repulsion by $U_{dd}=5.5$~eV and of
$p$--$d$ of the excited Co by $U_{\underline pd}=7.0$~eV; the
Slater integrals have been reduced to 80\% of their Hartree-Fock
value. An exchange field of $H_{\rm ex}=3$~meV has been used. For
the Co$_{\rm trig}$ ions the ionic crystal splittings are $\Delta
E^{\rm ionic}_{d_1/d_2}=0.9$~eV and $\Delta E^{\rm
ionic}_{d_0/d_2}=0.05$~eV taken from band-structure
calculation\cite{Wu:2005}; the hybridization
 is
 $V^{\rm hyb}_{d_1}=1.88$~eV,
 $V^{\rm hyb}_{d_0}=1.28$~eV, and
 $V^{\rm hyb}_{d_2}=1.25$~eV.
For Co$_{\rm oct}$ ions $pd\sigma=-1.44$~eV and
$pd\pi=0.63$~eV was used. With this set of parameters we have
found a LS--HS transition for Co$_{\rm oct}$ at $10Dq=0.65$~eV.
Here we used $10Dq=0.8$~eV, based on band-structure
calculation, which is the same value as for
EuCoO$_3$ known as a LS Co$^{3+}$ oxide.\cite{Hu:2004}

The results of these simulations are given by curves (b) in
Fig.~\ref{XMCD1}, together with a break-down into the separate
contributions of the Co$_{\rm trig}$ (curves c) and Co$_{\rm oct}$
(curves d) ions. One can clearly observe that the simulations
(curves b) reproduce the experimental spectra quite well. All
major and minor peaks in the individual experimental XAS ($\mu^+$,
$\mu^-$) and XMCD ($\mu^+-\mu^-$) spectra (curves a) are present
in the simulations (curves b). It is almost needless to remark
that the entire simulated XMCD signal is coming from the HS
Co$_{\rm trig}$ ions (curves c), since we started with a
nonmagnetic LS Co$_{\rm oct}$ (curves d). In the simulation, we
find $S_z = 1.8\mu_B$ and $L_z = 1.7\mu_B$, giving a total
magnetic moment of $2S_z +L_z = 5.3 \mu_B$ per formula unit.
Due to strong hybridization  between Co $3d$ and O $2p$,
the $3d$ occupation numbers of the Co$_{\rm trig}$ ions and the Co$_{\rm oct}$ are
6.3 and 6.8, respectively, giving on average 6.5
electrons as used for the sum rules. The calculated total moment from the simulation compares reasonably
well with the $4.8 \mu_B$ value from direct magnetization
measurements.\cite{Hardy:2004a,Maignan:2004} Yet, the simulated
$L_z$ value ($1.7\mu_B$) is appreciably larger than the one
derived from the experiment using the XMCD sum rule ($1.2\mu_B$).
However, looking more closely at the simulated and experimental
XMCD spectra, we can clearly see that the XMCD spectra have very
similar line shapes and that the distinction is mainly in the
amplitude, i.e. a matter of scaling. This indicates that the
discrepancy in the $L_z$ determination might
be caused by the fact that the sample is not
fully magnetized in our experiment. According to
magnetization measurements one can only achieve 90\% of the
saturation magnetization at 5.5~T.\cite{Maignan:2004}
In addition, slight misalignment of the sample together with the high magnetocrystalline anisotropy may account for some
further reduction of the experimental value.

\begin{figure}
    \includegraphics[width=0.5\textwidth]{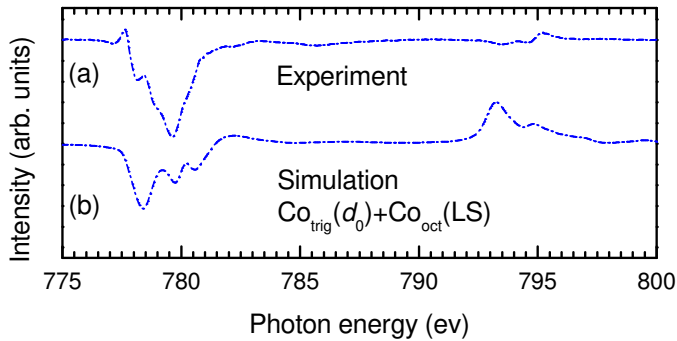}
    \caption{(Color online) (a) Measured soft-x-ray magnetic
     circular dichroism spectrum (XMCD, $\mu^+-\mu^-$);
     (b) Simulated XMCD spectrum assuming a doubly occupied $d_0$
     orbital for the Co$_{\rm trig}$ and low-spin (LS)
     Co$_{\rm oct}$ ions.}
    \label{XMCD2}
\end{figure}

Having established that the $d_2$/LS scenario for the Co$_{\rm
trig}$/Co$_{\rm oct}$ ions explains well the experimental
spectra, we now investigate the sensitivity of our analysis. For
this we change the \textit{ansatz} for the Co$_{\rm trig}$ ion: we
now put the $d_0$ orbital to be energetically lower than the
$d_2$.\cite{parametersD0} The result is shown in Fig.~\ref{XMCD2}, in which
the simulated XMCD spectrum (b) is compared with the experimental
one (a). One can unambiguously recognize large discrepancies in
the line shapes, not only in the $L_3$ region (777--785 eV), where
the simulated XMCD signal has much less amplitude, but also in
the $L_2$ (792--797 eV), where now an XMCD signal is calculated
while it is practically absent in the experiment. Using the XMCD
sum rule,\cite{Thole:1992} we can relate these discrepancies also
directly to the fact that the $d_0$ \textit{ansatz} essentially
does not carry an orbital moment.

\begin{figure}
    \includegraphics[width=0.5\textwidth]{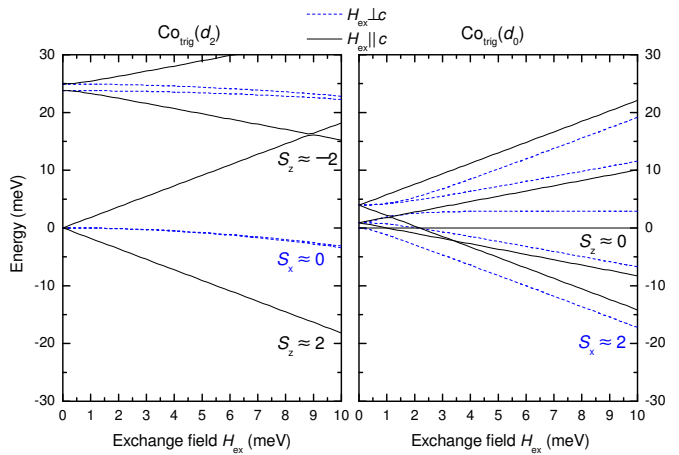}
    \caption{(Color online) Total-energy level diagram for
    the Co$_{\rm trig}$ ion as a function of the exchange field
    $H_{ex}$ along the $z$ direction ($c$ axis, solid black lines) and
    along the $x$ direction (perpendicular to the $c$ axis, dashed blue
    lines), with (left panel) the $d_2$ and (right panel)
    the $d_0$ orbital doubly occupied.}
    \label{energydiagram}
\end{figure}

The success of the cluster method for the analysis of both the
high-energy spectroscopies and the ground-state magnetic moments
provides confidence for its use to investigate the
magnetocrystalline anisotropy in this material. We have calculated
the total energy of the Co$_{\rm trig}$O$_6$ cluster as a function
of the exchange field $H_{\rm ex}$, directed either along the $z$ axis
(the $c$ axis or the one-dimensional chain) or along the $x$ axis
(perpendicular to the $c$ axis). The results for the $d_2$
\textit{ansatz} are plotted in the left panel of
Fig.~\ref{energydiagram}. It is evident that the lowest state gains
energy when $H_{\rm ex}$ is along $z$ and very little to nothing for
$H_{\rm ex}$ along $x$. This demonstrates directly that the
magnetocrystalline anisotropy should be large and directed along
the $z$, consistent with the experimental observation that
the magnetization parallel to $c$ is almost sixfold of the one
perpendicular $c$ (at 12~K and 5.5~T)\cite{Kageyama:1997b} and
with the Ising nature of the magnetism.
We note that the lowest curve in the figure shows an
energy gain with a rate twice that of $H_{\rm ex}$ (along $z$),
meaning that $S_z$ is close to $2\mu_B$ which in turn is
consistent with the HS ($S=2$) nature of the Co$_{\rm trig}$ ion.

We have also analyzed the magnetocrystalline anisotropy of the
Co$_{\rm trig}$O$_6$ cluster using the $d_0$ \textit{ansatz}. The
right panel of Fig. \ref{energydiagram} reveals that the total energy decreases for
both directions of $H_{\rm ex}$, but now with the important
distinction that the energy for $H_{\rm ex}$ along $x$ is always lower
than for $H_{\rm ex}$ along $z$. This implies that the easy-magnetization
axis should be perpendicular to the $z$ axis, which
is not consistent with the experimental facts. Also this contradiction
thus effectively falsifies the $d_0$ \textit{ansatz}.\cite{Whangbo:2003,Dai:2005}

To conclude, using soft x-ray absorption spectroscopy and the
magnetic circular dichroism therein at the Co-$L_{2,3}$ edges we
have experimentally determined that both Co ions in
Ca$_{3}$Co$_{2}$O$_{6}$ are 3+, with the Co$_{\rm trig}$ ion in
the high-spin state and the Co$_{\rm oct}$ in the nonmagnetic
state. The Co$_{\rm trig}$ ion carries an anomalously large
orbital moment of $1.7\mu_B$ which we have been able to relate to
the double occupation of the $d_2$ orbital. In addition, the
detailed analysis of the spectral line shapes together with that
of the magnetocrystalline anisotropy firmly establishes that the
$d_2$ orbital lies lowest in energy\cite{Wu:2005} and not the
$d_0$.\cite{Whangbo:2003,Dai:2005} This in turn also demonstrates
that a proper incorporation of the spin-orbit interaction is
required for the \textit{ab-initio} calculation of the delicate
electronic structure of this material.

We are grateful to Hua Wu and Daniel Khomskii for stimulating
discussions. The research in K\"oln is supported by the Deutsche
Forschungsgemeinschaft (DFG) through SFB 608.


\begin{thebibliography}{33}
\expandafter\ifx\csname natexlab\endcsname\relax\def\natexlab#1{#1}\fi
\expandafter\ifx\csname bibnamefont\endcsname\relax
  \def\bibnamefont#1{#1}\fi
\expandafter\ifx\csname bibfnamefont\endcsname\relax
  \def\bibfnamefont#1{#1}\fi
\expandafter\ifx\csname citenamefont\endcsname\relax
  \def\citenamefont#1{#1}\fi
\expandafter\ifx\csname url\endcsname\relax
  \def\url#1{\texttt{#1}}\fi
\expandafter\ifx\csname urlprefix\endcsname\relax\def\urlprefix{URL }\fi
\providecommand{\bibinfo}[2]{#2}
\providecommand{\eprint}[2][]{\url{#2}}

\bibitem[{\citenamefont{Aasland et~al.}(1997)\citenamefont{Aasland,
  Fjellv{\aa}g, and Hauback}}]{Aasland:1997}
\bibinfo{author}{\bibfnamefont{S.}~\bibnamefont{Aasland}},
  \bibinfo{author}{\bibfnamefont{H.}~\bibnamefont{Fjellv{\aa}g}},
  \bibnamefont{and} \bibinfo{author}{\bibfnamefont{B.}~\bibnamefont{Hauback}},
  \bibinfo{journal}{Solid State Commun.} \textbf{\bibinfo{volume}{101}},
  \bibinfo{pages}{187} (\bibinfo{year}{1997}).

\bibitem[{\citenamefont{Kageyama
  et~al.}(1997{\natexlab{a}})\citenamefont{Kageyama, Yoshimura, Kosuge,
  Mitamura, and Goto}}]{Kageyama:1997}
\bibinfo{author}{\bibfnamefont{H.}~\bibnamefont{Kageyama}},
  \bibinfo{author}{\bibfnamefont{K.}~\bibnamefont{Yoshimura}},
  \bibinfo{author}{\bibfnamefont{K.}~\bibnamefont{Kosuge}},
  \bibinfo{author}{\bibfnamefont{H.}~\bibnamefont{Mitamura}}, \bibnamefont{and}
  \bibinfo{author}{\bibfnamefont{T.}~\bibnamefont{Goto}}, \bibinfo{journal}{J.
  Phys. Soc. Jpn.} \textbf{\bibinfo{volume}{66}}, \bibinfo{pages}{1607}
  (\bibinfo{year}{1997}{\natexlab{a}}).

\bibitem[{\citenamefont{Kageyama
  et~al.}(1997{\natexlab{b}})\citenamefont{Kageyama, Yoshimura, Kosuge, Azuma,
  Takano, Mitamura, and Goto}}]{Kageyama:1997b}
\bibinfo{author}{\bibfnamefont{H.}~\bibnamefont{Kageyama}},
  \bibinfo{author}{\bibfnamefont{K.}~\bibnamefont{Yoshimura}},
  \bibinfo{author}{\bibfnamefont{K.}~\bibnamefont{Kosuge}},
  \bibinfo{author}{\bibfnamefont{M.}~\bibnamefont{Azuma}},
  \bibinfo{author}{\bibfnamefont{M.}~\bibnamefont{Takano}},
  \bibinfo{author}{\bibfnamefont{H.}~\bibnamefont{Mitamura}}, \bibnamefont{and}
  \bibinfo{author}{\bibfnamefont{T.}~\bibnamefont{Goto}}, \bibinfo{journal}{J.
  Phys. Soc. Jpn.} \textbf{\bibinfo{volume}{66}}, \bibinfo{pages}{3996}
  (\bibinfo{year}{1997}{\natexlab{b}}).

\bibitem[{\citenamefont{Maignan et~al.}(2000)\citenamefont{Maignan, Michel,
  Masset, Martin, and Raveau}}]{Maignan:2000}
\bibinfo{author}{\bibfnamefont{A.}~\bibnamefont{Maignan}},
  \bibinfo{author}{\bibfnamefont{C.}~\bibnamefont{Michel}},
  \bibinfo{author}{\bibfnamefont{A.~C.} \bibnamefont{Masset}},
  \bibinfo{author}{\bibfnamefont{C.}~\bibnamefont{Martin}}, \bibnamefont{and}
  \bibinfo{author}{\bibfnamefont{B.}~\bibnamefont{Raveau}},
  \bibinfo{journal}{Eur. Phys. J. B} \textbf{\bibinfo{volume}{15}},
  \bibinfo{pages}{657} (\bibinfo{year}{2000}).

\bibitem[{\citenamefont{Kudasov}(2006)}]{Kudasov:2006}
\bibinfo{author}{\bibfnamefont{Y.~B.} \bibnamefont{Kudasov}},
  \bibinfo{journal}{Phys. Rev. Lett.} \textbf{\bibinfo{volume}{96}},
  \bibinfo{pages}{027212} (\bibinfo{year}{2006}).

\bibitem[{\citenamefont{Hardy et~al.}(2004{\natexlab{a}})\citenamefont{Hardy,
  Lees, Petrenko, Paul, Flahaut, Hebert, and Maignan}}]{Hardy:2004a}
\bibinfo{author}{\bibfnamefont{V.}~\bibnamefont{Hardy}},
  \bibinfo{author}{\bibfnamefont{M.~R.} \bibnamefont{Lees}},
  \bibinfo{author}{\bibfnamefont{O.~A.} \bibnamefont{Petrenko}},
  \bibinfo{author}{\bibfnamefont{D.~McK.} \bibnamefont{Paul}},
  \bibinfo{author}{\bibfnamefont{D.}~\bibnamefont{Flahaut}},
  \bibinfo{author}{\bibfnamefont{S.}~\bibnamefont{Hebert}}, \bibnamefont{and}
  \bibinfo{author}{\bibfnamefont{A.}~\bibnamefont{Maignan}},
  \bibinfo{journal}{Phys. Rev. B} \textbf{\bibinfo{volume}{70}},
  \bibinfo{pages}{064424} (\bibinfo{year}{2004}{\natexlab{a}}).

\bibitem[{\citenamefont{Flahaut et~al.}(2004)\citenamefont{Flahaut, Maignan,
  H\'ebert, Martin, Retoux, and Hardy}}]{Flahaut:2004}
\bibinfo{author}{\bibfnamefont{D.}~\bibnamefont{Flahaut}},
  \bibinfo{author}{\bibfnamefont{A.}~\bibnamefont{Maignan}},
  \bibinfo{author}{\bibfnamefont{S.}~\bibnamefont{H\'ebert}},
  \bibinfo{author}{\bibfnamefont{C.}~\bibnamefont{Martin}},
  \bibinfo{author}{\bibfnamefont{R.}~\bibnamefont{Retoux}}, \bibnamefont{and}
  \bibinfo{author}{\bibfnamefont{V.}~\bibnamefont{Hardy}},
  \bibinfo{journal}{Phys. Rev. B} \textbf{\bibinfo{volume}{70}},
  \bibinfo{pages}{094418} (\bibinfo{year}{2004}).

\bibitem[{\citenamefont{Hardy et~al.}(2004{\natexlab{b}})\citenamefont{Hardy,
  Flahaut, Lees, and Petrenko}}]{Hardy:2004}
\bibinfo{author}{\bibfnamefont{V.}~\bibnamefont{Hardy}},
  \bibinfo{author}{\bibfnamefont{D.}~\bibnamefont{Flahaut}},
  \bibinfo{author}{\bibfnamefont{M.~R.} \bibnamefont{Lees}}, \bibnamefont{and}
  \bibinfo{author}{\bibfnamefont{O.~A.} \bibnamefont{Petrenko}},
  \bibinfo{journal}{Phys. Rev. B} \textbf{\bibinfo{volume}{70}},
  \bibinfo{pages}{214439} (\bibinfo{year}{2004}{\natexlab{b}}).

\bibitem[{\citenamefont{Petrenko et~al.}(2005)\citenamefont{Petrenko,
  Wooldridge, Lees, Manuel, and Hardy}}]{Petrenko2005}
\bibinfo{author}{\bibfnamefont{O.}~\bibnamefont{Petrenko}},
  \bibinfo{author}{\bibfnamefont{J.}~\bibnamefont{Wooldridge}},
  \bibinfo{author}{\bibfnamefont{M.}~\bibnamefont{Lees}},
  \bibinfo{author}{\bibfnamefont{P.}~\bibnamefont{Manuel}}, \bibnamefont{and}
  \bibinfo{author}{\bibfnamefont{V.}~\bibnamefont{Hardy}},
  \bibinfo{journal}{Eur. Phys. J. B} \textbf{\bibinfo{volume}{47}},
  \bibinfo{pages}{79} (\bibinfo{year}{2005}).

\bibitem[{\citenamefont{Fjellv{\aa}g et~al.}(1996)\citenamefont{Fjellv{\aa}g,
  Gulbrandsen, Aasland, Olsen, and Hauback}}]{Fjellvaag:1996}
\bibinfo{author}{\bibfnamefont{H.}~\bibnamefont{Fjellv{\aa}g}},
  \bibinfo{author}{\bibfnamefont{E.}~\bibnamefont{Gulbrandsen}},
  \bibinfo{author}{\bibfnamefont{S.}~\bibnamefont{Aasland}},
  \bibinfo{author}{\bibfnamefont{A.}~\bibnamefont{Olsen}}, \bibnamefont{and}
  \bibinfo{author}{\bibfnamefont{B.~C.} \bibnamefont{Hauback}},
  \bibinfo{journal}{J. Solid State Chem.} \textbf{\bibinfo{volume}{124}},
  \bibinfo{pages}{190} (\bibinfo{year}{1996}).

\bibitem[{\citenamefont{Hardy et~al.}(2003)\citenamefont{Hardy, Lambert, Lees,
  and Paul}}]{Hardy:2003}
\bibinfo{author}{\bibfnamefont{V.}~\bibnamefont{Hardy}},
  \bibinfo{author}{\bibfnamefont{S.}~\bibnamefont{Lambert}},
  \bibinfo{author}{\bibfnamefont{M.~R.} \bibnamefont{Lees}}, \bibnamefont{and}
  \bibinfo{author}{\bibfnamefont{D.~McK.} \bibnamefont{Paul}},
  \bibinfo{journal}{Phys. Rev. B} \textbf{\bibinfo{volume}{68}},
  \bibinfo{pages}{014424} (\bibinfo{year}{2003}).

\bibitem[{\citenamefont{Maignan et~al.}(2004)\citenamefont{Maignan, Hardy,
  Hebert, Drillon, Lees, Petrenko, Paul, and Khomskii}}]{Maignan:2004}
\bibinfo{author}{\bibfnamefont{A.}~\bibnamefont{Maignan}},
  \bibinfo{author}{\bibfnamefont{V.}~\bibnamefont{Hardy}},
  \bibinfo{author}{\bibfnamefont{S.}~\bibnamefont{Hebert}},
  \bibinfo{author}{\bibfnamefont{M.}~\bibnamefont{Drillon}},
  \bibinfo{author}{\bibfnamefont{M.}~\bibnamefont{Lees}},
  \bibinfo{author}{\bibfnamefont{O.}~\bibnamefont{Petrenko}},
  \bibinfo{author}{\bibfnamefont{D.~McK.} \bibnamefont{Paul}},
  \bibnamefont{and} \bibinfo{author}{\bibfnamefont{D.}~\bibnamefont{Khomskii}},
  \bibinfo{journal}{J. Mater. Chem.} \textbf{\bibinfo{volume}{14}},
  \bibinfo{pages}{1231} (\bibinfo{year}{2004}).

\bibitem[{\citenamefont{Vidya et~al.}(2003)\citenamefont{Vidya, Ravindran,
  Fjellv{\aa}g, Kjekshus, and Eriksson}}]{Vidya:2003}
\bibinfo{author}{\bibfnamefont{R.}~\bibnamefont{Vidya}},
  \bibinfo{author}{\bibfnamefont{P.}~\bibnamefont{Ravindran}},
  \bibinfo{author}{\bibfnamefont{H.}~\bibnamefont{Fjellv{\aa}g}},
  \bibinfo{author}{\bibfnamefont{A.}~\bibnamefont{Kjekshus}}, \bibnamefont{and}
  \bibinfo{author}{\bibfnamefont{O.}~\bibnamefont{Eriksson}},
  \bibinfo{journal}{Phys. Rev. Lett.} \textbf{\bibinfo{volume}{91}},
  \bibinfo{eid}{186404} (\bibinfo{year}{2003}).

\bibitem[{\citenamefont{Sampathkumaran
  et~al.}(2004)\citenamefont{Sampathkumaran, Fujiwara, Rayaprol, Madhu, and
  Uwatoko}}]{Sampathkumaran:2004}
\bibinfo{author}{\bibfnamefont{E.~V.} \bibnamefont{Sampathkumaran}},
  \bibinfo{author}{\bibfnamefont{N.}~\bibnamefont{Fujiwara}},
  \bibinfo{author}{\bibfnamefont{S.}~\bibnamefont{Rayaprol}},
  \bibinfo{author}{\bibfnamefont{P.~K.} \bibnamefont{Madhu}}, \bibnamefont{and}
  \bibinfo{author}{\bibfnamefont{Y.}~\bibnamefont{Uwatoko}},
  \bibinfo{journal}{Phys. Rev. B} \textbf{\bibinfo{volume}{70}},
  \bibinfo{pages}{014437} (\bibinfo{year}{2004}).

\bibitem[{\citenamefont{Takubo et~al.}(2005)\citenamefont{Takubo, Mizokawa,
  Hirata, Son, Fujimori, Topwal, Sarma, Rayaprol, and
  Sampathkumaran}}]{Takubo:2005}
\bibinfo{author}{\bibfnamefont{K.}~\bibnamefont{Takubo}},
  \bibinfo{author}{\bibfnamefont{T.}~\bibnamefont{Mizokawa}},
  \bibinfo{author}{\bibfnamefont{S.}~\bibnamefont{Hirata}},
  \bibinfo{author}{\bibfnamefont{J.-Y.} \bibnamefont{Son}},
  \bibinfo{author}{\bibfnamefont{A.}~\bibnamefont{Fujimori}},
  \bibinfo{author}{\bibfnamefont{D.}~\bibnamefont{Topwal}},
  \bibinfo{author}{\bibfnamefont{D.~D.} \bibnamefont{Sarma}},
  \bibinfo{author}{\bibfnamefont{S.}~\bibnamefont{Rayaprol}}, \bibnamefont{and}
  \bibinfo{author}{\bibfnamefont{E.-V.} \bibnamefont{Sampathkumaran}},
  \bibinfo{journal}{Phys. Rev. B} \textbf{\bibinfo{volume}{71}},
  \bibinfo{pages}{073406} (\bibinfo{year}{2005}).

\bibitem[{\citenamefont{Eyert et~al.}(2004)\citenamefont{Eyert, Laschinger,
  Kopp, and Fr\'esard}}]{Eyert:2004}
\bibinfo{author}{\bibfnamefont{V.}~\bibnamefont{Eyert}},
  \bibinfo{author}{\bibfnamefont{C.}~\bibnamefont{Laschinger}},
  \bibinfo{author}{\bibfnamefont{T.}~\bibnamefont{Kopp}}, \bibnamefont{and}
  \bibinfo{author}{\bibfnamefont{R.}~\bibnamefont{Fr\'esard}},
  \bibinfo{journal}{Chem. Phys. Lett.} \textbf{\bibinfo{volume}{385}},
  \bibinfo{pages}{249} (\bibinfo{year}{2004}).

\bibitem[{\citenamefont{Whangbo et~al.}(2003)\citenamefont{Whangbo, Dai, Koo,
  and Jobic}}]{Whangbo:2003}
\bibinfo{author}{\bibfnamefont{M.-H.} \bibnamefont{Whangbo}},
  \bibinfo{author}{\bibfnamefont{D.}~\bibnamefont{Dai}},
  \bibinfo{author}{\bibfnamefont{H.-J.} \bibnamefont{Koo}}, \bibnamefont{and}
  \bibinfo{author}{\bibfnamefont{S.}~\bibnamefont{Jobic}},
  \bibinfo{journal}{Solid State Commun.} \textbf{\bibinfo{volume}{125}},
  \bibinfo{pages}{413} (\bibinfo{year}{2003}).

\bibitem[{\citenamefont{{Hua Wu} et~al.}(2005)\citenamefont{{Hua Wu},
  Haverkort, Hu, Khomskii, and Tjeng}}]{Wu:2005}
\bibinfo{author}{\bibnamefont{{Hua Wu}}}, \bibinfo{author}{\bibfnamefont{M.~W.}
  \bibnamefont{Haverkort}},
  \bibinfo{author}{\bibfnamefont{Z.}~\bibnamefont{Hu}},
  \bibinfo{author}{\bibfnamefont{D.~I.} \bibnamefont{Khomskii}},
  \bibnamefont{and} \bibinfo{author}{\bibfnamefont{L.~H.} \bibnamefont{Tjeng}},
  \bibinfo{journal}{Phys. Rev. Lett.} \textbf{\bibinfo{volume}{95}},
  \bibinfo{pages}{186401} (\bibinfo{year}{2005}).

\bibitem[{\citenamefont{Dai and Whangbo}(2005)}]{Dai:2005}
\bibinfo{author}{\bibfnamefont{D.}~\bibnamefont{Dai}} \bibnamefont{and}
  \bibinfo{author}{\bibfnamefont{M.-H.} \bibnamefont{Whangbo}},
  \bibinfo{journal}{Inorg. Chem.} \textbf{\bibinfo{volume}{44}},
  \bibinfo{pages}{4407} (\bibinfo{year}{2005}).

\bibitem[{\citenamefont{Hu et~al.}(2004)\citenamefont{Hu, Wu, Haverkort, Hsieh,
  Lin, Lorenz, Baier, Reichl, Bonn, Felser et~al.}}]{Hu:2004}
\bibinfo{author}{\bibfnamefont{Z.}~\bibnamefont{Hu}},
  \bibinfo{author}{\bibfnamefont{H.}~\bibnamefont{Wu}},
  \bibinfo{author}{\bibfnamefont{M.~W.} \bibnamefont{Haverkort}},
  \bibinfo{author}{\bibfnamefont{H.~H.} \bibnamefont{Hsieh}},
  \bibinfo{author}{\bibfnamefont{H.~J.} \bibnamefont{Lin}},
  \bibinfo{author}{\bibfnamefont{T.}~\bibnamefont{Lorenz}},
  \bibinfo{author}{\bibfnamefont{J.}~\bibnamefont{Baier}},
  \bibinfo{author}{\bibfnamefont{A.}~\bibnamefont{Reichl}},
  \bibinfo{author}{\bibfnamefont{I.}~\bibnamefont{Bonn}},
  \bibinfo{author}{\bibfnamefont{C.}~\bibnamefont{Felser}},
  \bibnamefont{et~al.}, \bibinfo{journal}{Phys. Rev. Lett.}
  \textbf{\bibinfo{volume}{92}}, \bibinfo{pages}{207402}
  (\bibinfo{year}{2004}).

\bibitem[{\citenamefont{Mitra et~al.}(2003)\citenamefont{Mitra, Hu,
  Raychaudhuri, Wirth, Csiszar, Hsieh, Lin, Chen, and Tjeng}}]{Mitra2003}
\bibinfo{author}{\bibfnamefont{C.}~\bibnamefont{Mitra}},
  \bibinfo{author}{\bibfnamefont{Z.}~\bibnamefont{Hu}},
  \bibinfo{author}{\bibfnamefont{P.}~\bibnamefont{Raychaudhuri}},
  \bibinfo{author}{\bibfnamefont{S.}~\bibnamefont{Wirth}},
  \bibinfo{author}{\bibfnamefont{S.~I.} \bibnamefont{Csiszar}},
  \bibinfo{author}{\bibfnamefont{H.~H.} \bibnamefont{Hsieh}},
  \bibinfo{author}{\bibfnamefont{H.-J.} \bibnamefont{Lin}},
  \bibinfo{author}{\bibfnamefont{C.~T.} \bibnamefont{Chen}}, \bibnamefont{and}
  \bibinfo{author}{\bibfnamefont{L.~H.} \bibnamefont{Tjeng}},
  \bibinfo{journal}{Phys. Rev. B} \textbf{\bibinfo{volume}{67}},
  \bibinfo{eid}{092404} (\bibinfo{year}{2003}).

\bibitem[{\citenamefont{Thole et~al.}(1992)\citenamefont{Thole, Carra, Sette,
  and van~der Laan}}]{Thole:1992}
\bibinfo{author}{\bibfnamefont{B.~T.} \bibnamefont{Thole}},
  \bibinfo{author}{\bibfnamefont{P.}~\bibnamefont{Carra}},
  \bibinfo{author}{\bibfnamefont{F.}~\bibnamefont{Sette}}, \bibnamefont{and}
  \bibinfo{author}{\bibfnamefont{G.}~\bibnamefont{van~der Laan}},
  \bibinfo{journal}{Phys. Rev. Lett.} \textbf{\bibinfo{volume}{68}},
  \bibinfo{pages}{1943} (\bibinfo{year}{1992}).

\bibitem[{\citenamefont{Saitoh et~al.}(1997)\citenamefont{Saitoh, Mizokawa,
  Fujimori, Abbate, Takeda, and Takano}}]{Saitoh1997}
\bibinfo{author}{\bibfnamefont{T.}~\bibnamefont{Saitoh}},
  \bibinfo{author}{\bibfnamefont{T.}~\bibnamefont{Mizokawa}},
  \bibinfo{author}{\bibfnamefont{A.}~\bibnamefont{Fujimori}},
  \bibinfo{author}{\bibfnamefont{M.}~\bibnamefont{Abbate}},
  \bibinfo{author}{\bibfnamefont{Y.}~\bibnamefont{Takeda}}, \bibnamefont{and}
  \bibinfo{author}{\bibfnamefont{M.}~\bibnamefont{Takano}},
  \bibinfo{journal}{Phys. Rev. B} \textbf{\bibinfo{volume}{55}},
  \bibinfo{pages}{4257} (\bibinfo{year}{1997}).

\bibitem[{\citenamefont{Tanaka and Jo}(2004)}]{Tanaka94}
\bibinfo{author}{\bibfnamefont{A.}~\bibnamefont{Tanaka}} \bibnamefont{and}
  \bibinfo{author}{\bibfnamefont{T.}~\bibnamefont{Jo}}, \bibinfo{journal}{J.
  Phys. Soc. Jpn.} \textbf{\bibinfo{volume}{63}}, \bibinfo{pages}{2788}
  (\bibinfo{year}{2004}).

\bibitem[{\citenamefont{de~Groot}(1994)}]{deGroot94}
\bibinfo{author}{\bibfnamefont{F.~M.~F.} \bibnamefont{de~Groot}},
  \bibinfo{journal}{J. Electron Spectrosc. Rel. Phenom.}
  \textbf{\bibinfo{volume}{67}}, \bibinfo{pages}{529} (\bibinfo{year}{1994}).

\bibitem[{\citenamefont{{See the ``Theo Thole Memorial
  Issue"}}(1997)}]{Thole97}
\bibinfo{author}{\bibnamefont{{See the ``Theo Thole Memorial Issue"}}},
  \bibinfo{journal}{J. Electron Spectrosc. Rel. Phenom.}
  \textbf{\bibinfo{volume}{86}}, \bibinfo{pages}{1} (\bibinfo{year}{1997}).

\bibitem[{\citenamefont{Saitoh et~al.}(1995)\citenamefont{Saitoh, Bocquet,
  Mizokawa, and Fujimori}}]{Saitoh1995}
\bibinfo{author}{\bibfnamefont{T.}~\bibnamefont{Saitoh}},
  \bibinfo{author}{\bibfnamefont{A.~E.} \bibnamefont{Bocquet}},
  \bibinfo{author}{\bibfnamefont{T.}~\bibnamefont{Mizokawa}}, \bibnamefont{and}
  \bibinfo{author}{\bibfnamefont{A.}~\bibnamefont{Fujimori}},
  \bibinfo{journal}{Phys. Rev. B} \textbf{\bibinfo{volume}{52}},
  \bibinfo{pages}{7934} (\bibinfo{year}{1995}).

\bibitem[{\citenamefont{Hu et~al.}(1998)\citenamefont{Hu, Mazumdar, Kaindl,
  de~Groot, Warda, and Reinen}}]{Hu:1998}
\bibinfo{author}{\bibfnamefont{Z.}~\bibnamefont{Hu}},
  \bibinfo{author}{\bibfnamefont{C.}~\bibnamefont{Mazumdar}},
  \bibinfo{author}{\bibfnamefont{G.}~\bibnamefont{Kaindl}},
  \bibinfo{author}{\bibfnamefont{F.~M.~F.} \bibnamefont{de~Groot}},
  \bibinfo{author}{\bibfnamefont{S.~A.} \bibnamefont{Warda}}, \bibnamefont{and}
  \bibinfo{author}{\bibfnamefont{D.}~\bibnamefont{Reinen}},
  \bibinfo{journal}{Chem. Phys. Lett.} \textbf{\bibinfo{volume}{297}},
  \bibinfo{pages}{321} (\bibinfo{year}{1998}).

\bibitem[{\citenamefont{Goodenough}(1965)}]{Goodenough:1971}
\bibinfo{author}{\bibfnamefont{J.~B.} \bibnamefont{Goodenough}}, in
  \emph{\bibinfo{booktitle}{Progress in Solid State Chemistry, Vol. 5}}, edited
  by \bibinfo{editor}{\bibfnamefont{H.}~\bibnamefont{Reiss}}
  (\bibinfo{publisher}{Pergamon Press}, \bibinfo{address}{Oxford},
  \bibinfo{year}{1965}), p. \bibinfo{pages}{145}.

\bibitem[{\citenamefont{Radaelli and
  Cheong}(2002{\natexlab{b}})}]{Radaelli2002}
\bibinfo{author}{\bibfnamefont{P.~G.} \bibnamefont{Radaelli}} \bibnamefont{and}
  \bibinfo{author}{\bibfnamefont{S.-W.} \bibnamefont{Cheong}},
  \bibinfo{journal}{Phys. Rev. B} \textbf{\bibinfo{volume}{66}},
  \bibinfo{pages}{094408} (\bibinfo{year}{2002}{\natexlab{b}}).

\bibitem[{\citenamefont{Bocquet et~al.}(1996)\citenamefont{Bocquet, Mizokawa,
  Morikawa, Fujimori, Barman, Maiti, Sarma, Tokura, and Onoda}}]{Bocquet96}
\bibinfo{author}{\bibfnamefont{A.~E.} \bibnamefont{Bocquet}},
  \bibinfo{author}{\bibfnamefont{T.}~\bibnamefont{Mizokawa}},
  \bibinfo{author}{\bibfnamefont{K.}~\bibnamefont{Morikawa}},
  \bibinfo{author}{\bibfnamefont{A.}~\bibnamefont{Fujimori}},
  \bibinfo{author}{\bibfnamefont{S.~R.} \bibnamefont{Barman}},
  \bibinfo{author}{\bibfnamefont{K.}~\bibnamefont{Maiti}},
  \bibinfo{author}{\bibfnamefont{D.~D.} \bibnamefont{Sarma}},
  \bibinfo{author}{\bibfnamefont{Y.}~\bibnamefont{Tokura}}, \bibnamefont{and}
  \bibinfo{author}{\bibfnamefont{M.}~\bibnamefont{Onoda}},
  \bibinfo{journal}{Phys. Rev. B} \textbf{\bibinfo{volume}{53}},
  \bibinfo{pages}{1161} (\bibinfo{year}{1996}).

\bibitem[{par()}]{parametersD0}
\bibinfo{note}{$\Delta E^{\rm total}_{d_2/d_0}=0.102\rm~eV$ with an exchange
  field of $H_{\rm ex}=3\rm~meV$, $H_{\rm ex}||c$.}

\end{thebibliography}
\end{document}